\documentclass[10pt, twocolumn, journal]{IEEEtran}

\usepackage{comment}
\usepackage{amsmath,amssymb,amsthm,amsfonts}
\usepackage{mathtools}
\usepackage{bigints}
\usepackage{url}
\usepackage[printonlyused]{acronym}
\usepackage{nameref}
\usepackage{prettyref}
\usepackage{nicefrac}
\usepackage{afterpage}
\usepackage[sort,nocompress]{cite}
\usepackage[normalem]{ulem}
\usepackage{soul}
\usepackage[draft, silent]{fixme}
\usepackage{caption}
\usepackage{array}
\usepackage{graphicx}
\usepackage{authblk}
\graphicspath{{./figures/}}

\newcommand{\gammabar}{\bar{\gamma}}

\newcommand{\M}{\mathcal{M}}

\newcommand{\Kd}{\mathcal{K}}

\newcommand{\wepsilon}{w^{\varepsilon}}

\def\P{{\mathbb{P}}}  
\def\minplus{{$(\min, +)\,$}} 
\def\S{{\cal S}}
\def\A{{\cal A}}
\def\D{{\cal D}}

\def\K{{\cal K}}
\def\C{{\cal C}}

\def\M{{\mathcal M}}

\def\Per{{\mathrm {P}}}

\def\mx{{$(\min, \times)$}}

\def\L{\mathbb{L}}



\begin{document}


\title{Statistical Delay Bound for WirelessHART Networks}

\author{
    \IEEEauthorblockN{Neda Petreska\IEEEauthorrefmark{1}, Hussein Al-Zubaidy\IEEEauthorrefmark{4}, Barbara Staehle\IEEEauthorrefmark{5}, Rudi Knorr\IEEEauthorrefmark{1}, James Gross\IEEEauthorrefmark{4}} \\
    \IEEEauthorblockA{\IEEEauthorrefmark{1}Fraunhofer Institute for Embedded Systems and Communication Technologies ESK
    \\\{neda.petreska, rudi.knorr\}@esk.fraunhofer.de} \\
    \IEEEauthorblockA{\IEEEauthorrefmark{4}School of Electrical Engineering, KTH Royal Institute of Technology
    \\{hzubaidy@kth.se, james.gross@ee.kth.se}
}\\
    \IEEEauthorblockA{\IEEEauthorrefmark{5}Faculty of Computer Science, University of Applied Sciences Konstanz
    \\{bstaehle@htwg-konstanz.de}
}}
\renewcommand\Authands{, }
\maketitle
\begin{abstract}
In this paper we provide a performance analysis framework for wireless industrial networks by deriving a service curve and a bound on the delay violation probability. For this purpose we use the \mx~stochastic network calculus as well as a recently presented recursive formula for an end-to-end delay bound of wireless heterogeneous networks. The derived results are mapped to WirelessHART networks used in process automation and were validated via simulations. In addition to WirelessHART, our results can be applied to any wireless network whose physical layer conforms the IEEE 802.15.4 standard, while its MAC protocol incorporates TDMA and channel hopping, like e.g. ISA100.11a or TSCH-based networks. The provided delay analysis is especially useful during the network design phase, offering further research potential towards optimal routing and power management in QoS-constrained wireless industrial networks.
\end{abstract}


\section{Motivation}
\label{sec:mot}

During the last decade we have witnessed an increasing usage of wireless networks in industrial settings. Due to their flexibility and low maintenance cost, a growing number of industrial applications have been realized with wireless instead of wired networks, e.g. in chemical, construction, automotive and agriculture industry, covering broad spectrum of process and factory automation scenarios. While factory automation tends to have strict QoS requirements, such as target delays smaller than 10 ms, usually even less than 1 ms and outage probabilities smaller than $10^{-6}$, process automation applications have looser latency demands, usually in the order of hundreds of milliseconds and delay violation probabilities not bigger than $10^{-3}$ \cite{zvei, zvei2}. Low data rates, typically not greater than a couple of kbps, characterize the mentioned cases of automation applications.

Due to their relatively looser QoS demands in comparison to factory automation, process automation applications are more often implemented by wireless communication technologies in practice. Typical applications in the area of process automation are predictive maintenance, control and monitoring applications, often realized by sensor networks \cite{jounela}. Several wireless technologies suitable for process automation, such as WirelessHART \cite{whart_online}, \mbox{ISA100.11a~\cite{isa100.11a}} and Industrial WLAN~\cite{iwlan}, have emerged in the recent years. WirelessHART and ISA100.11a, both based on the IEEE 802.15.4-2006 PHY standard~\cite{802_15_4_standard}, offer multi-hop communication, especially useful for energy-limited sensor networks when larger distances between two nodes have to be bridged. 
Since applications are characterized by end-to-end QoS requirements, typically expressed with the delay and its violation probability, performance analysis of wireless networks is necessary and contributes to a reliable and QoS-aware network behaviour, particularly important for network design and flow admission estimation.

On the other hand, the random nature of the wireless channel contributes to a random instantaneous channel capacity, resulting into outages and an unreliable network behaviour. As a result, buffers are built into the wireless transceivers: a behaviour that has to be taken into account, due to its strong influence on the packet delay. All these facts increase the complexity of performance analysis of wireless networks in general. 

Motivated by the above mentioned, in this paper we present a closed-form expression for the statistical delay bound as a part of performance analysis framework for wireless industrial networks, whose physical layer is defined according to the IEEE 802.15.4 standard, being one of the most frequently used standards for wireless industrial networks. In particular, we derive a service curve of a wireless link within a WirelessHART network, where we use the bit error rate (BER) definition as a function of the signal-to-noice ratio (SNR) given in the mentioned standard. We use the obtained service curve together with a definition on the arrival process to define the bound on the stochastic delay, i.e., the delay violation probability. Using previous results on the end-to-end delay bound for heterogeneous wireless networks \cite{icc15} based on the stochastic network calculus, we enable performance analysis of wireless multi-hop industrial networks under statistical delay requirements. 
Certain modifications on the MAC layer defined in~\cite{802_15_4_standard}, such as in WirelessHART, ISA100.11a or TSCH networks (the last one defined within the IEEE 802.15.4e standard~\cite{802_15_4e_standard}), characterized by both TDMA and frequency-hopping, enable the usage of our network calculus framework on a broad family of technologies. 
To the best of our knowledge, this is the first attempt to determine a service curve and its resulting delay bound modeling precisely the system under investigation instead of opting for the usual approach of using the Shannon capacity model. The channel capacity overestimation resulting from the Shannon model is highlighted in the numerical section of the paper. 

The paper is structured as follows: Section~\ref{sec:relwork} discusses related work. The system model, problem statement, an introduction to WirelessHART and the theoretical framework of stochastic network calculus are given in Section~\ref{sec:preliminaries}. The main contribution of the paper, i.e., the derivation of the WirelessHART service curve and delay bound is presented in Section~\ref{sec:sc}. The analytical expressions are validated in Section~\ref{sec:results}, where also additional numerical results are presented. Section~\ref{sec:conclusion} concludes the paper.

\section{Related Work}
\label{sec:relwork}

Many works address performance evaluation of 802.15.4-based networks, often opting for simulation and/or real test-bed evaluation and measurements \cite{ferrari_whart, song, lee}. However, there are not many works presenting analytical approaches for network performance of industrial wireless systems.
\cite{whart_perf_eval} models WirelessHART networks' performance using a Discrete-Time Markov Chain, considering link's SNR and BER, and a path model, predicting path performance and providing routing suggestions. Using this model, the authors compute reachability, link availability, delay and utilization, however, without capturing any queuing behaviour. 
The authors of \cite{saifullah_journal} derive an upper bound on the end-to-end delays of flows under fixed priority scheduling in WirelessHART networks. They map the real-time transmission scheduling to real-time multiprocessor scheduling and use its methods for the delay bound formulation. 
The derived results are validated through simulations for different WirelessHART scenarios and topologies.  An analytical Markov model that predicts the performance and detailed behavior of the 802.15.4 slotted Carrier Sense Multiple Access/Collision Avoidance (CSMA/CA) mechanism is given in~\cite{pollin}. The Markov model is used to capture the state of each user at each moment of time and to define various performance parameters, like e.g. the probability of starting a new transmission attempt following a successful or failed packet transmission. The model is validated via simulations and the analysis predicts the energy consumption and throughput in 802.15.4 networks. 

In comparison to the discussed works, ~\cite{zhu} and~\cite{schmitt} incorporate queuing effects into the delay analysis. \cite{zhu} proposes a stochastic analysis approach to evaluate the delay performance of a CSMA/CA scheme for a one-hop beacon-enabled 802.15.4 network. They combine Discrete Time Markov Chains and M/G/1/K queues of finite buffered nodes to derive expressions for the channel access probability, the busy channel probability, such as the probability distribution of packet queue size. However, the analysis of average packet service delays is not suitable in case of traffic bursts and wireless fading channels. Finally,~\cite{schmitt} presents the so called sensor network calculus as a tool for worst case traffic analysis in sensor networks. The authors discuss which service curves are suitable so that sensor characteristics, such as duty cycle or energy consumption, are integrated. However, using deterministic bounds for performance guarantees of wireless networks leaves further potential for improvements. 
  
\section{Preliminaries}
\label{sec:preliminaries}

In this section we introduce the WirelessHART technology and then present the considered system model. A short background on \mx~stochastic network calculus, used for the derivation of the analytical results in Section~\ref{sec:sc}, is provided at the end of the section.
\subsection{Introduction to WirelessHART}
\label{subsec:whart}

\begin{figure}
\centering
  \includegraphics[scale=0.5]{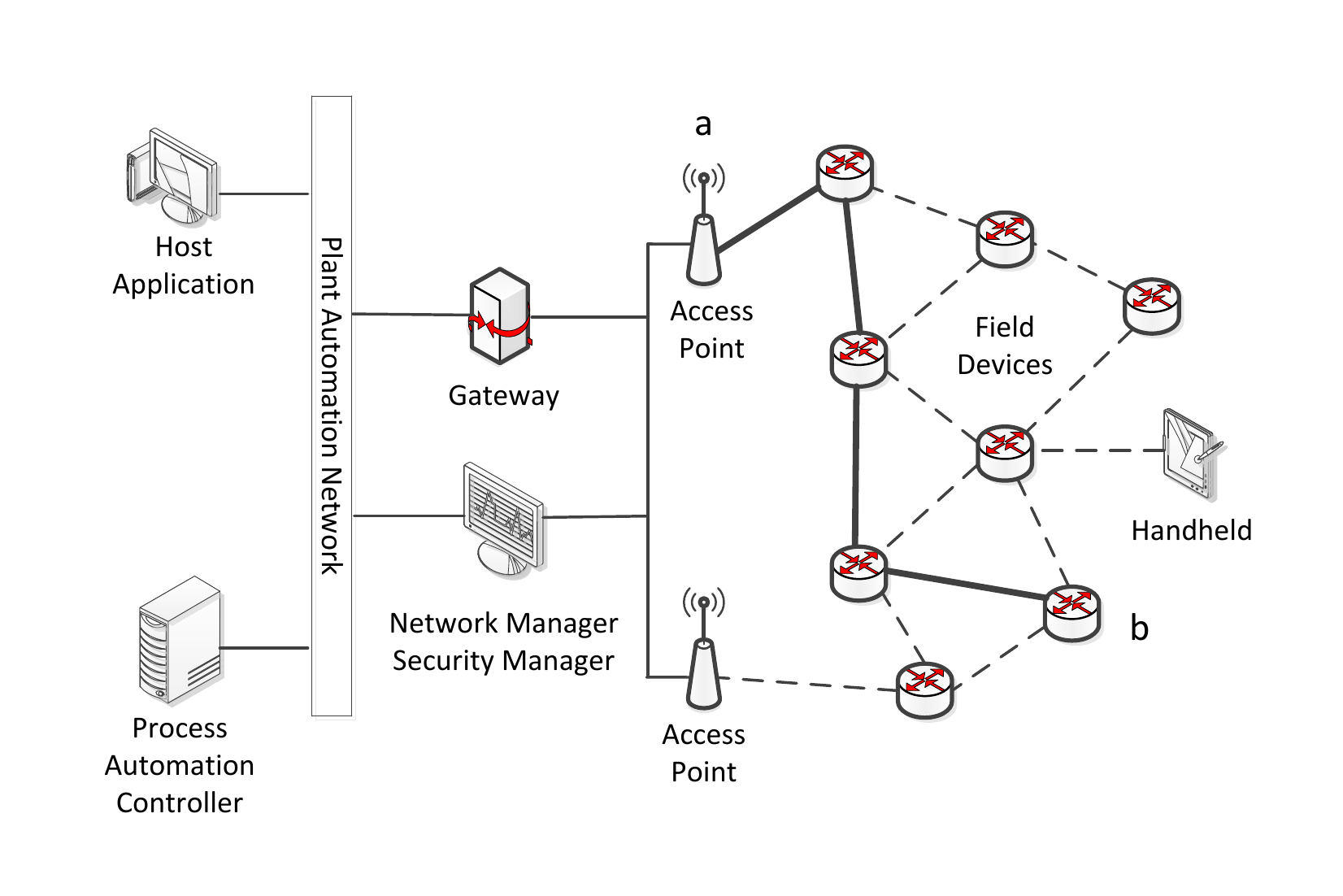}
\caption{WirelessHART network architecture \cite{whart_architecture}}
\label{fig:whart_arch}
\end{figure}

The WirelessHART technology \cite{whart_online} is the wireless alternative of the Highway Addressable Remote Tranducer Protocol (HART) and is mainly intended for non-time-critical process automation applications with battery-limited devices. The network is centrally controlled by a so called \textit{network manager}, which assigns transmission resources to communication nodes referred to as \textit{field devices}. The generic WirelessHART network architecture is presented in Fig.~\ref{fig:whart_arch}. 
The physical layer is based on the IEEE 802.15.4 standard for low power networks, while on the medium access and network layer the Time Synchronized Mesh Protocol (TSMP)~\cite{tsmp} is applied. 
In TSMP the radio-frequency space is modeled as a matrix of slot-channel \emph{cells}. 
Each cell in the matrix lasts for one \emph{time slot of 10 ms}. The cells are grouped in so called \emph{superframes}, which repeat at a constant rate. 
TSMP changes the transmission channel on a per-slot basis, i.e., it employs frequency hopping. This increases the bandwidth, reduces multi-path fading effects and is more robust to interference, while at the same time reducing the impact on other neighbouring wireless networks. 
A collision-free system operation can be guaranteed if each event (data transmission) is scheduled in a new cell, therefore in a different time slot and on a different frequency. 
In order to achieve that, the assignment of cells per node is done by the network manager, using the control channel. 
According to the TSMP time slot format, more than half of each slot is overhead (acknowledgment (ACK) frame and synchronization preambles). 
About $4$ ms (or 250 symbols) remain for payload transmission. 

According to the IEEE 802.15.4 standard, a sequence of consecutive and equally sized time slots builds a superframe. A superframe can last from 15 ms up to several seconds and its format is defined by the network manager. Within every superframe, the network manager allocates a specific number of time slots to each device, during which the data is being transmitted. There are 15 channels used for data transmission and one control channel.
The maximal frame size, as defined in IEEE 802.15.4, is 133 bytes, out of which 6 bytes belong to the header. A 16-bit cyclic redundancy code (CRC) is used for error detection, leading to the whole frame being dropped in case of a bit error.

\subsection{System Model and Problem Statement}
\label{subsec:sysmodel}

We observe a communication flow originating at a source node $a$ and ending at a destination node $b$, traversing an $n$-hop path within a WirelessHART network\footnote{Note that usually the communication in WirelessHART takes place between the gateway and the field devices. We focus, however, in this paper exclusively on multi-hop wireless communication.} (see solid thick lines in Fig.~\ref{fig:whart_arch}). Let this multi-hop path be given by \mbox{$\L=\{1,..,n\}$}. Each node contains a buffer where the packets can be queued up in case of an unsuccessful transmission, i.e., no reception of an ACK. The system is time-slotted and the central network manager allocates one cell (a combination of time slot and sending frequency) per superframe for each node. 
Let $T$ be the length of a time slot (in our case $T=10$ ms) and $N$ the number of time slots within a superframe. 
Hence, a superframe lasts for exactly $T\cdot N$ time units. 
In our system model, we observe one multi-hop path consisting of as many links as there are time slots per superframe, i.e., $n=N$ links. We further assume a round-robin link scheduling fashion: as soon as the last link finishes with the transmission, the next superframe begins, the first link is scheduled again and so on. To simplify, the network manager always assigns the $j$-th time slot within one superframe to the $j$-th link along the path, while the channel changes in a random fashion. 
We assume block-fading channels, where the instantaneous SNR of link $j$, $\gamma_{i,j}$, remains constant within its designated time slot in superframe $i$ and changes from one superframe to another. We denote its average with $\gammabar_{j}$. 
Two consecutive transmissions by the same node (and on the same link) are exactly $N$ time slots apart, so the instantaneous SNRs of the same link are independent and uncorrelated to each other. 


At the beginning of each superframe, the application layer of the source node generates a payload of $r_a$ bits, stored at its sending buffer. 
The maximal number of bits that can be transmitted in a WirelessHART slot is set to $k_a = 1016$ bits. Hence, depending on $r_a$, several packets can be transmitted within one data frame. 
Because of the CRC, the instantaneous wireless service of link $j$ in superframe $i$, $s_{i,j}$, is a random variable and we define it as $s_{i,j} \equiv X_{i,j}$. $X_{i,j}$ can take the values of either 0 or $k_a$, following the Bernoulli distribution. The application has QoS requirements given with the pair $\{\wepsilon, \varepsilon\}$, where $\wepsilon$ is a so called statistical delay and represents the maximal delay that the application packets can experience, while $\varepsilon$ is the maximal tolerable probability with which $\wepsilon$ can be violated. 

In this work, we are interested in obtaining an analytical expression for the end-to-end delay bound for the depicted multi-hop path between nodes $a$ and $b$ in Fig.~\ref{fig:whart_arch}. In order to achieve this, we first need a characterization of the service offered by the path to the incoming data flow, originating from a process automation application running on node $a$. For the analysis we use the stochastic network calculus theory, shortly presented in the following subsection.

\subsection{Stochastic \mx~ Network Calculus}

Stochastic network calculus considers queuing systems and networks of systems with stochastic arrival and departure processes, where the bivariate functions $A(\tau,t)$, $D(\tau,t)$ and $S(\tau,t)$ for any $0 \le \tau \le t$, denote the \textit{cumulative} arrivals to the system, departures from the system, and service offered by the system, respectively, in the interval $[\tau,t)$.  
Recall that we consider a discrete time model, where time slots (or in our case superframes) have a duration $T$ and $i \geq 0$ denotes the index of the respective superframe.

A lossless system with an arrival process $A(\tau,t)$ and a service process $S(\tau,t)$ satisfies the relationship \mbox{$D(\tau,t) \geq A \otimes S \left(\tau,t\right)$}, where $\otimes$ is the \minplus~convolution operator given by
\begin{equation}
 x \otimes y \left(\tau,t\right) = \inf_{\tau \leq u \leq t} \left\{ x(\tau,u) + y (u,t) \right\} \; .
 \label{eq:convolution}
 \end{equation}
\noindent As stated above, in general we are interested in probabilistic  bounds of the form $\mathrm{Pr}\left[ W(t) > w^{\varepsilon} \right] \leq \varepsilon$, known as the \textit{violation probability} for a target delay $w^{\varepsilon}$, under stable system conditions:
\begin{equation}
\label{eq:stability}
\lim_{t\to \infty} {\frac{A(0,t)}{t}} < \lim_{t\to \infty} {\frac{S(0,t)}{t}}.
\end{equation}

Modeling wireless links in the context of network calculus however is not a trivial task. 
As in the case of effective capacity \cite{wu}, it is especially difficult to obtain a stochastic characterization of the cumulative service process of a wireless fading channel. 
A recent work \cite{alzubaidy} proposes that performance guarantees of wireless buffered and fading channels are expressed in a so called "SNR" domain, instead of the usually used bit domain \cite{jiang:servermodel, mahmood:mimo, fidler_mgf, wu}.
This can be interpreted as the \textit{SNR domain} (thinking of bits as "SNR demands" that reside in the system until they can be met by the channel).

The cumulative arrival, service, and departure processes in the bit domain, i.e., $A$, $D$, and $S$, are related to their SNR domain counterparts (represented in the following by calligraphic capital letters $\A$, $\D$, and $\S$) respectively, through the exponential function.
Thus, we have $\mathcal{A}(\tau,t) \triangleq e^{A(\tau,t)}$, $\mathcal{D}(\tau,t) \triangleq e^{D(\tau,t)}$, and $\mathcal{S}(\tau,t) \triangleq e^{S(\tau,t)}$.
Due to the exponential function, these cumulative processes become products of the increments in the bit domain. In the following, we will assume $\mathcal{A}\left(\tau,t\right)$ and $\mathcal{S}\left(\tau,t\right)$ to have stationary and independent increments.
We denote them by $\alpha$ for the arrivals (in SNR domain) and $g\left(\gamma\right)$ for the service. Hence, for the instantaneous service of a link in the $i$-th superframe we write $s_i=g(\gamma_i)$, where $\gamma_i$ is the instantaneous link's SNR and the cumulative service process in the SNR domain equals to
\begin{equation}
\label{eq:service_process_snr}
  \mathcal{S}(\tau,t)  = \prod_{i=\tau}^{t-1}  e^{s_i} = \prod_{i=\tau}^{t-1} g\left(\gamma_i\right).
\end{equation}
Furthermore, the delay at time $t$ is obtained
as follows:
\begin{equation}
\label{eq:delay_snr}
  W(t)=\mathcal{W}(t)=\inf \{i \geq 0 : \A(0,t) / \D(0,t+i) \leq 1 \}.
\end{equation}
A bound $\varepsilon$ for the delay violation probability $\mathrm{Pr}\left[W(t)>w^\varepsilon\right]$ can be derived based on a transform of the cumulative arrival and service process in the SNR domain using the moment bound.
In~\cite{alzubaidy} it was shown that such a violation probability bound for a given $\wepsilon$ can be obtained as
$\inf\limits_{s>0}\left\{\Kd(s, t+\wepsilon,t)\right\}$.
\noindent We refer to the function $\Kd\left(s,\tau,t\right)$ as the \textit{kernel} defined as
 \begin{equation}
\label{eq:function_M_Hussein}
  \Kd(s,-\wepsilon) = \sum_{i=0}^{\mathrm{min}(\tau,t)} \mathcal{M}_{\mathcal{A}}(1+s,i,t) \mathcal{M}_{\mathcal{S}}(1-s,i,\tau),
\end{equation}
where the function $\mathcal{M}_{\mathcal{X}}\left(s\right)$ is the Mellin transform \cite{Book:mellin} of a random process, defined as
\begin{equation}
\mathcal{M}_{\mathcal{X}}\left(s,\tau,t\right) = \mathcal{M}_{\mathcal{X}\left(\tau,t\right)} \left(s\right) = \mathbb{E}\left[\mathcal{X}^{s-1}\left(\tau,t\right)\right],
\label{eq:Mellin_Definition}
\end{equation}
for any $s \in \mathbb{C}$ (we restrict our derivations in this work to real values $s \in \mathbb{R}$). 
Introducing the Mellin transform in the performance analysis of wireless fading channels enables easier mathematical manipulation when performing network calculus bounds computation as well as scalable closed-form solutions. For stationary processes the Mellin transforms become independent of the time instance and we write $\mathcal{M}_{\mathcal{X}}\left(s,\tau,t\right) = \mathcal{M}_{\mathcal{X}}\left(s, t - \tau\right)$.
%
In addition, as we only consider stable queuing systems in steady-state, the kernel becomes independent of the time instance $t$ and we denote $\Kd\left(s,t+\wepsilon,t \right) \overset{t \to \infty}{=} \Kd\left(s,-\wepsilon\right)$.

The strength of the Mellin-transform-based approach becomes apparent when considering block-fading channels.
The Mellin transform for the cumulative service process in the SNR domain is given by
\begin{equation}
\label{eq:mellin_transform_service_basic}
 \mathcal{M}_{\mathcal{S}}\left(s,\tau,t\right)=\prod_{i=\tau}^{t-1} \mathcal{M}_{g(\gamma)}\left(s\right)=\mathcal{M}_{g(\gamma)}^{t-\tau}\left(s\right)= \mathcal{M}_{\S}\left(s,t-\tau\right)\, 
\end{equation}
where $\mathcal{M}_{g(\gamma)}\left(s\right)$ is the Mellin transform of the stationary and independent service increment $g\left(\gamma\right)$ in the SNR domain.
The function $g\left(\cdot\right)$ represents here the channel capacity. However, it can also model more complex system characteristics, most importantly scheduling effects.

Assuming the cumulative arrival process in the SNR domain to have stationary and independent increments we denote the corresponding Mellin transform by 
\begin{equation}
\M_{\A}\left(s,t - \tau\right) = \prod_{i=\tau}^{t-1}\M_{\alpha}(s) = \M_{\alpha}^{t-\tau}(s).
\end{equation} 
Substituting these two cumulative processes in Eq.~\eqref{eq:function_M_Hussein}, for the steady-state kernel of a fading wireless channel we get~\cite{itc14}
\begin{equation}
  \Kd\left(s,-\wepsilon\right) = \frac{\left(\M_{g\left(\gamma\right)}\left(1 - s\right)\right)^w}{1 - \M_{\alpha}\left(1 + s \right) \M_{g\left(\gamma\right)}\left(1 - s \right)}
  \label{eq:delay_kernel}
\end{equation}
for any $s > 0$, under the stability condition
\begin{equation}
\label{eq:stability_cond}
  \M_{\alpha}\left(1 + s \right) \M_{g\left(\gamma\right)}\left(1 - s \right) < 1.
\end{equation}

\section{W\MakeLowercase{ireless}HART Delay Bound}
\label{sec:sc}

In this section we first present the derivation of the kernel for a WirelessHART system. We then shortly discuss the application area of our proposed performance analysis framework beyond a  WirelessHART network setup. 

\subsection{Delay Bound Derivation}
\label{subsec:derivation}

Since a MAC-frame in WirelessHART is dropped as soon as any of its $k_a$ bits has been erroneously transmitted, the frame error rate (FER) of link $j$ in superframe $i$ is given by 
\begin{equation}
\label{eq:per}
\Per(\gamma_{i,j})=1-(1-p(\gamma_{i,j}))^{k_a}.
\end{equation}
The BER $p(\gamma_{i,j})$ is a function of the instantaneous SNR $\gamma_{i,j}$ on a particular link $j$ in the $i$-th superframe and is given by the IEEE 802.15.4 standard \cite{802_15_4_standard}:
\begin{equation}
\label{eq:BER_802154}
  p(\gamma_{i,j})=\frac{1}{30} \sum_{u=2}^{16} (-1)^u \binom{16}{u} e^{-20\gamma_{i,j}(1-\nicefrac{1}{u})}.
\end{equation}
%
The random service $X_{i,j}$ offered by the $j$-th link to the data flow in the $i$-th superframe is Bernoulli distributed, since either the whole frame is successfully transmitted or it is completely dropped:
\begin{equation}
\label{eq:xi}
  X_{i,j}=
  \begin{cases}
     k_a, &  1-\Per (\gamma_{i,j}) \\
     0, &  \Per (\gamma_{i,j})
  \end{cases}
\end{equation}
\noindent We represent the cumulative service of link $j$ in the bit domain in the time interval $(\tau,t)$ as $S_j(\tau,t) = \sum_{i=\tau}^{t-1} X_{i,j}$. The cumulative service in the SNR domain results in
\begin{equation}
\label{eq:service_snr_v3}
  \S_j(\tau,t) = e^{S_j(\tau,t)} = e^{\sum_{i=\tau}^{t-1} X_{i,j}} = \prod_{i=\tau}^{t-1} e^{X_{i,j}}
\end{equation}
\noindent and its Mellin transform is computed as follows:
\begin{equation}
\label{eq:mellin_service_v3}
\begin{aligned}
  \M_{\S_j(\tau,t)}(s) & = \mathbb{E} \left[(\S_j(\tau,t))^{s-1}\right] = \mathbb{E} \left[\left(\prod_{i=\tau}^{t-1}e^{X_{i,j}}\right)^{s-1} \right] = \\
  & = \prod_{i=\tau}^{t-1} \mathbb{E} \left[ \left(e^{X_{i,j}}\right)^{(s-1)}\right] = \left[\M_{e^{X_{i,j}}}(s)\right]^{(t-\tau)}.
\end{aligned}
\end{equation}
\noindent $\M_{e^{X_{i,j}}}(s), s>0$, represents the Mellin transform of the service of the $j$-th link in its assigned time slot for transmission within the $i$-th superframe. According to our system model, between two consecutive transmissions on the same link within an $n$-hop path lay exactly $n$ time slots or $n\cdot 10$ ms. Furthermore, according to TSMP, each time the link is newly assigned a different transmission channel is used. These two facts lead to independent $X_{i,j}$ events, which in turn enable the product of the expectations in Eq.~\eqref{eq:mellin_service_v3}. Since the service is Bernoulli distributed, we write:
\begin{equation}
\label{eq:mellin_timeslot_2}
\begin{aligned}
  & \M_{e^{X_{i,j}}}(s)  = \mathbb{E} \left[e^{X_{i,j}(s-1)} \right] = \\
  & = e^{k_a(s-1)} \cdot \P(X_{i,j}=k_a) + e^{0} \cdot \P(X_{i,j}=0). 
\end{aligned}
\end{equation}
\noindent We define $\P(X_{i,j}=x)$ as follows:
\begin{equation}
\label{eq:bernouli_pmf}
\P(X_{i,j}=x)=
\begin{cases}
1-\Per(\gamma_{i,j})=(1-p(\gamma_{i,j}))^{k_a}, & x=k_a, \\
\Per(\gamma_{i,j})=1-(1-p(\gamma_{i,j}))^{k_a}, & x=0
\end{cases}
\end{equation}
\noindent and also as a marginal probability distribution:
\begin{equation}
\label{eq:marginal}
\P(X_{i,j}=x)=\int_0^{\infty} \P(X_{i,j}=x|\gamma_{i,j}=y)\cdot \P(\gamma_{i,j}=y)dy
\end{equation}
Combining Eq.~\eqref{eq:bernouli_pmf} and Eq.~\eqref{eq:marginal} and assuming a Rayleigh-fading channel with exponentially distributed SNR with mean $\gammabar_j$, we obtain
\begin{equation}
\label{eq:bernouli_pmf2}
\P(X_{i,j}=x)=
\begin{cases}
\int_{0}^{\infty} (1-p(y))^{k_a}\cdot \frac{1}{\gammabar_j}e^{\nicefrac{-y}{\gammabar_j}}dy, & x=k_a, \\
\int_{0}^{\infty} (1-(1-p(y))^{k_a})\cdot \frac{1}{\gammabar_j}e^{\nicefrac{-y}{\gammabar_j}}dy, & x=0.
\end{cases}
\end{equation}
Substituting Eq.~\eqref{eq:bernouli_pmf2} in Eq.~\eqref{eq:mellin_timeslot_2} we get
\begin{equation}
\label{eq:mellin_xij}
\begin{aligned}
\M_{e^{X_{i,j}}}(s) & = e^{k_a(s-1)} \cdot \int_{0}^{\infty} (1-p(y))^{k_a}\cdot \frac{1}{\gammabar_j}e^{\nicefrac{-y}{\gammabar_j}}dy \\
& + \int_{0}^{\infty} (1-(1-p(y))^{k_a})\cdot \frac{1}{\gammabar_j}e^{\nicefrac{-y}{\gammabar_j}}dy.
\end{aligned}
\end{equation}
Because of the complex form of $p(\gamma_{i,j})$ (see Eq.~\eqref{eq:BER_802154}), Eq.~\eqref{eq:mellin_xij} can be best solved numerically. We provide however, in the following, the analytical form that $\M_{\S_j(\tau,t)}(s)$ takes.

We start by solving the integral $\int_{0}^{\infty} (1-p(y))^{k_a}\cdot \frac{1}{\gammabar_j}e^{\nicefrac{-y}{\gammabar_j}}dy$. Here, we use the binomial theorem $(a+b)^n=\sum_{k=0}^n\binom{n}{k} a^{n-k}b^k$ and obtain:
\begin{align}
\label{eq:int_1}
& \P(X_{i,j}=k_a)= \int_{0}^{\infty} (1-p(y))^{k_a}\cdot \frac{1}{\gammabar_j}e^{\nicefrac{-y}{\gammabar_j}}dy \nonumber \\ 
& =\frac{1}{\gammabar_j} \sum_{k=0}^{k_a} \int_{0}^{\infty} \binom{k_a}{k}(-p(y))^{k}e^{\nicefrac{-y}{\gammabar_j}}dy = \frac{1}{\gammabar_j} \sum_{k=0}^{k_a} \nonumber \\
& \int_{0}^{\infty} \binom{k_a}{k}\left(-\frac{1}{30} \sum_{u=2}^{16} (-1)^u \binom{16}{u} e^{-20y(1-\nicefrac{1}{u})}\right)^{k}e^{\nicefrac{-y}{\gammabar_j}}dy \nonumber \\ 
& =  \frac{1}{\gammabar_j} \sum_{k=0}^{k_a} \int_{0}^{\infty} \left(\binom{k_a}{k} \sum_{r=1}^{c_k} A_{r,k}\cdot e^{-B_{r,k}\cdot y-\frac{y}{\gammabar_j}}\right)dy \nonumber \\
& = \sum_{k=0}^{k_a} \binom{k_a}{k} \sum_{r=1}^{c_k} \frac{A_{r,k}}{B_{r,k}+\frac{1}{\gammabar_j}},
\end{align}
\noindent where the coefficients $A_{r,k}$ and $B_{r,k}$ result from the power calculation of the expression \\ \mbox{$\left(-\frac{1}{30} \sum_{u=2}^{16} (-1)^u \binom{16}{u}e^{-20y(1-\nicefrac{1}{u})}\right)^{k}$} and therefore depend on $k$. 
$c_k$ defines the number of summands that result from this power calculation.

Similarly, for the second marginal probability we obtain:
\begin{equation}
\label{eq:int_2}
\begin{aligned}
\P(X_{i,j}=0) & =\int_{0}^{\infty} (1-(1-p(y))^{k_a})\cdot \frac{1}{\gammabar_j}e^{\nicefrac{-y}{\gammabar_j}}dy \\
& = 1-\sum_{k=0}^{k_a} \binom{k_a}{k} \sum_{r=1}^{c_k} \frac{A_{r,k}}{B_{r,k}+\frac{1}{\gammabar_j}}.
\end{aligned}
\end{equation}

Substituting Eq.~\eqref{eq:int_1} and Eq.~\eqref{eq:int_2} into Eq.~\eqref{eq:mellin_xij}, we obtain $\M_{e^{X_{i,j}}}(s)$. Having this, as well as substituting \\ \mbox{$Q(\gammabar_j)=\sum_{k=0}^{k_a} \binom{k_a}{k} \sum_{r=1}^{c_k} \frac{A_{r,k}}{B_{r,k}+\frac{1}{\gammabar_j}}$}, we get the following expression for the Mellin transform of the cumulative service in a WirelessHART system: 
\begin{align}
\M_{\S_j(\tau,t)}(s)=\left(1+ (e^{k_a(s-1)}-1)Q(\gammabar_j)\right)^{(t-\tau)}=\left(\M_{\beta_j}(s)\right)^{t-\tau}.
\end{align}

In order to obtain the kernel, we need to consider the arrival process as well. The Mellin transform of the arrival in the SNR domain, \mbox{$\A(\tau,t)=e^{r_a(t-\tau)}$}, is given by $\M_{\A}(s,\tau,t) = e^{r_a(t-\tau)(s-1)}$. 
According to Eq.~\eqref{eq:function_M_Hussein}, the delay bound for a given target probabilistic delay $w$ in case of a single-hop communication results in:
\begin{equation}
  \K^{\{j\}}(s,t+w,t) = \sum_{i=0}^{t} e^{r_a(t-i)s}\left(1+ (e^{-k_as}-1)Q(\gammabar_j)\right)^{t+w-i}.
\end{equation}
For a stable system, we let $t \rightarrow \infty$ and do change of variables $t-i=v$, so we obtain:
\begin{equation}
\label{eq:func_M}
\begin{aligned}
  & \K^{\{j\}}(s,-w) = \sum_{v=0}^{\infty} e^{r_avs}\left(1+ (e^{-k_as}-1)Q(\gammabar_j)\right)^{v+w}= \\
  & = \left(1+ (e^{-k_as}-1)Q(\gammabar_j)\right)^{w} \sum_{v=0}^{\infty} \left(e^{r_as}\left(1+ (e^{-k_as}-1)Q(\gammabar_j)\right)\right)^v \\
	& =\frac{\left(1+ (e^{-k_as}-1)Q(\gammabar_j)\right)^{w} }{1-e^{r_as}\left(1+ (e^{-k_as}-1)Q(\gammabar_j)\right)}.
\end{aligned}
\end{equation}
The sum converges if the stability condition
\begin{equation}
\label{eq:stability_whart}
\begin{aligned}
  & e^{r_as}(1+ (e^{-k_as}-1)Q(\gammabar_j)) < 1 \\
  \Leftrightarrow & r_a < -\frac{1}{s}\log{(1+ (e^{-k_as}-1)Q(\gammabar_j))}, \\
\end{aligned}
\end{equation}
is met. The end-to-end delay bound $\K^{\L}(s,-w)$ of the given multi-hop path $\L$ is computed using the following recursive formula (for any $m \in \{1,..,n-1\}$, Theorem 1 in \cite{icc15}):
\begin{equation}
\label{eq:recursion}
\begin{aligned}
 \K^{\L}(s,-w) & =\frac{\M_{\beta_n}(1-s)}{\M_{\beta_n}(1-s)-\M_{\beta_m}(1-s)}\K^{\L\setminus \{m\}}(s,-w) \\
& + \frac{\M_{\beta_m}(1-s)}{\M_{\beta_m}(1-s)-\M_{\beta_n}(1-s)}\K^{\L\setminus \{n\}}(s,-w).
\end{aligned}
\end{equation}
\noindent 
Eq.~\eqref{eq:func_M} is substituted as the basic case of the recursion. 

\subsection{Application Area of the Delay Bound}
\label{subsec:app_area}

The usage of the BER expression defined in the IEEE 802.15.4 standard extends the application area of the derived results beyond the WirelessHART standard. More specifically, the obtained results can be transferred to any technology which meets the following two conditions: (1) its physical layer description is based on the standard~\cite{802_15_4_standard} and (2) its MAC layer implementation provides TDMA and channel hopping. The second condition, which enables the computation of the Mellin transform of the service as defined in Eq.~\eqref{eq:mellin_service_v3}, for which independent service increments are necessary, is fulfilled by a time-slotted and frequency hopping MAC scheme. Hence, our results apply also in ISA100.11a networks as well as networks based on the IEEE 802.15.4e standard~\cite{802_15_4e_standard}. The latter one enhances the MAC protocol of IEEE 802.15.4 by introducing the Time Synchronized Channel Hopping (TSCH) MAC behaviour mode with dedicated links in order to increase network capacity and reliability.

\section{Numerical Results}
\label{sec:results}

In this section we present the validation of the derived analytical results as well as numerical examples discussing the influence of parameters, such as the SNR, the hop number and the payload size on the WirelessHART kernel\footnote{Note that the following discussion applies also to the other technologies mentioned in Sec.~\ref{subsec:app_area}.}.

\subsection{Methodology}
We have simulated several multi-hop paths, in which each link is characterized by a different average SNR and an unlimited buffer size. Each iteration represents one superframe during which every link in a round-robin fashion forwards the received payload along the path. The packet of link $j$ in superframe $i$ is dropped with certain probability $\Per (\gamma_{i,j})$ as defined in  Sec.~\ref{subsec:derivation}. At the beginning of each superframe new $r_a$ bits enter the system. Also, an instantaneous SNR per link is drawn from an exponential distribution with pre-defined mean value. Afterward, the BER per link is computed as given by Eq.~\eqref{eq:BER_802154}, followed by determining the FER $\Per (\gamma_{i,j})$ according to Eq.~\eqref{eq:per}. The instantaneous service per link in each superframe is drawn from the Bernoulli distribution (Eq.~\eqref{eq:xi}). The time is marked as soon as the payload enters the first node on the path and leaves the destination node. At the end, the number of packets with delay larger than the target one is determined. For the computation of the analytical end-to-end delay bound we use Eq.~\eqref{eq:recursion}. We set the frame size to $k_a=1016$ bits, as defined in~\cite{802_15_4_standard}.

\subsection{Validation and Discussion}
We now present the validation of the analytical delay bound together with results showing the influence of certain parameters on the QoS performance.
\begin{figure}
\centering
  \includegraphics[scale=0.6]{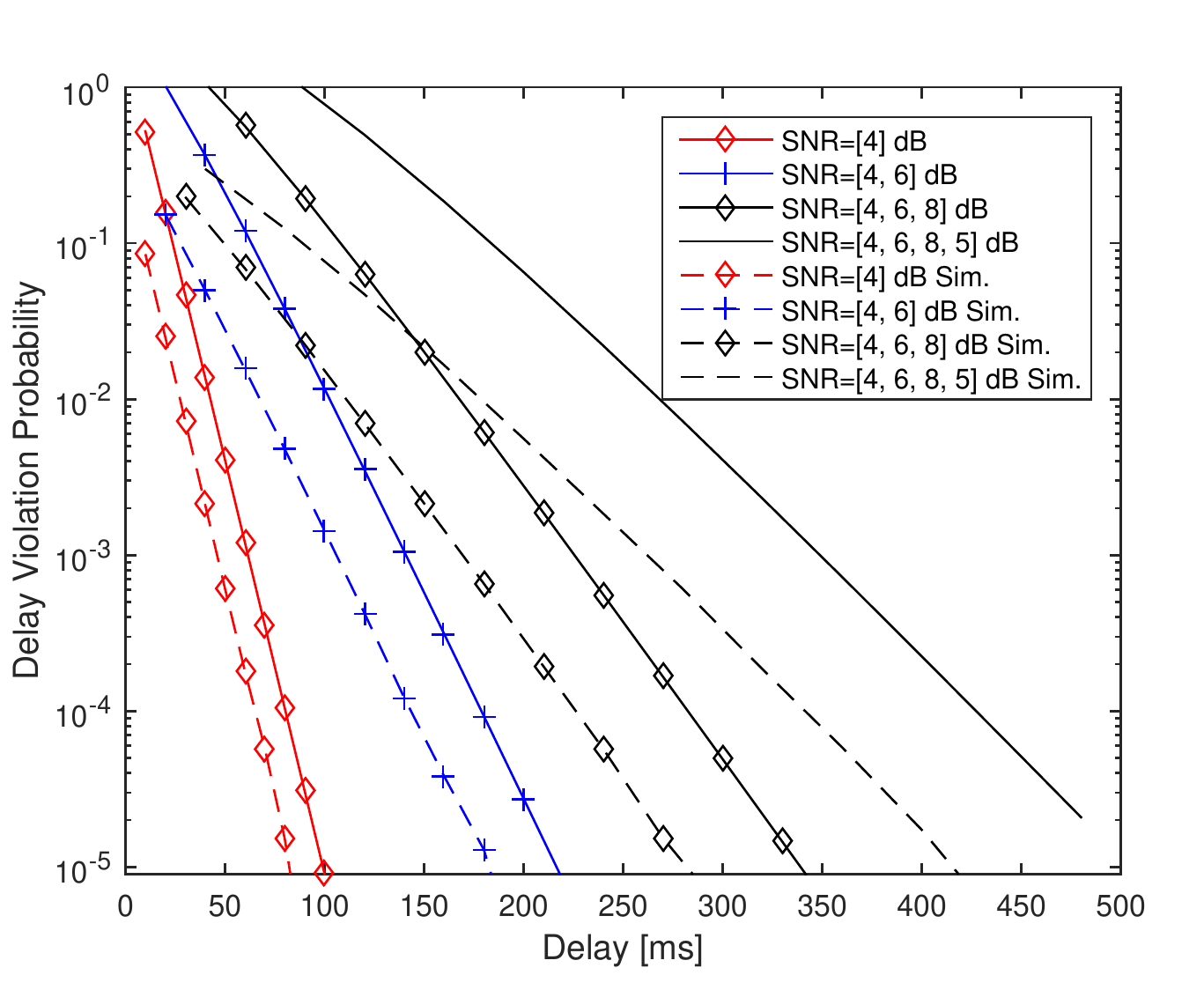}
\caption{Validation of the delay bound for a WirelessHART system for different multi-hop scenarios.}
\label{fig:validation_whart}
\end{figure}

Fig.~\ref{fig:validation_whart} shows the analytical delay bound in comparison to the empirical delay violation likelihood for different target delays and different path compositions with 1 to 4 hops characterized with various average SNR. 
The simulated delay bound is represented with dashed lines. 
We notice that the analytical delay bound given with Eq.~\eqref{eq:func_M} and Eq.~\eqref{eq:recursion} is an upper bound of the simulated kernel. 
The tightness of the bound depends on the number of hops, i.e., for longer paths, the bound gets less tight, but even for 4 hops, the gap between the simulated and the analytical delay bound is still one order of magnitude.	

\begin{figure}[ht]
  \includegraphics[scale=0.6]{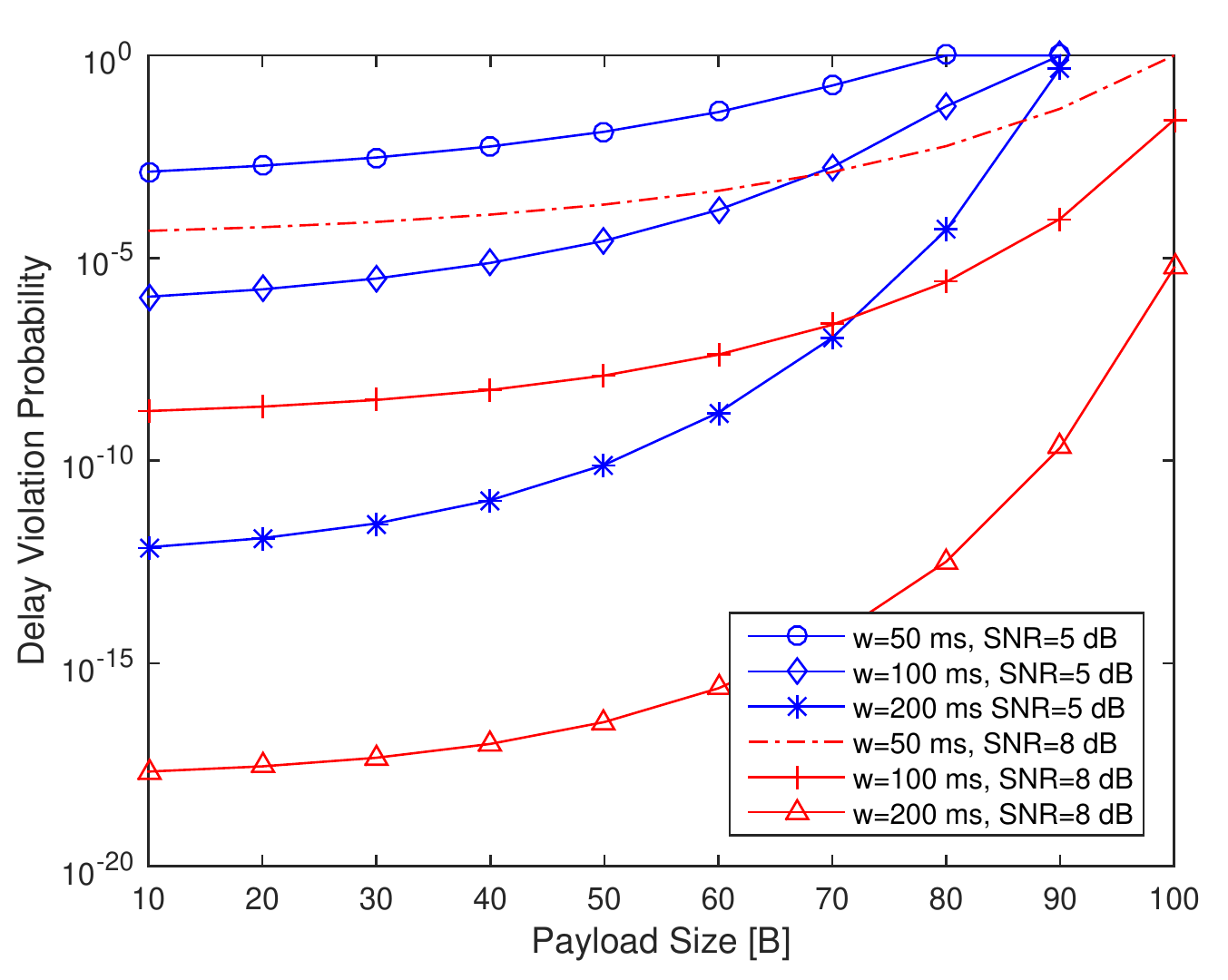} 
\caption{Delay bound computed depending on the payload size $r_a$ for different target delays and in case when $\gammabar \in \{5, 8\}$ dB in a single-hop scenario.}
\label{fig:diff_ra}
\end{figure}
\begin{figure}[ht]
\centering
  \includegraphics[scale=0.6]{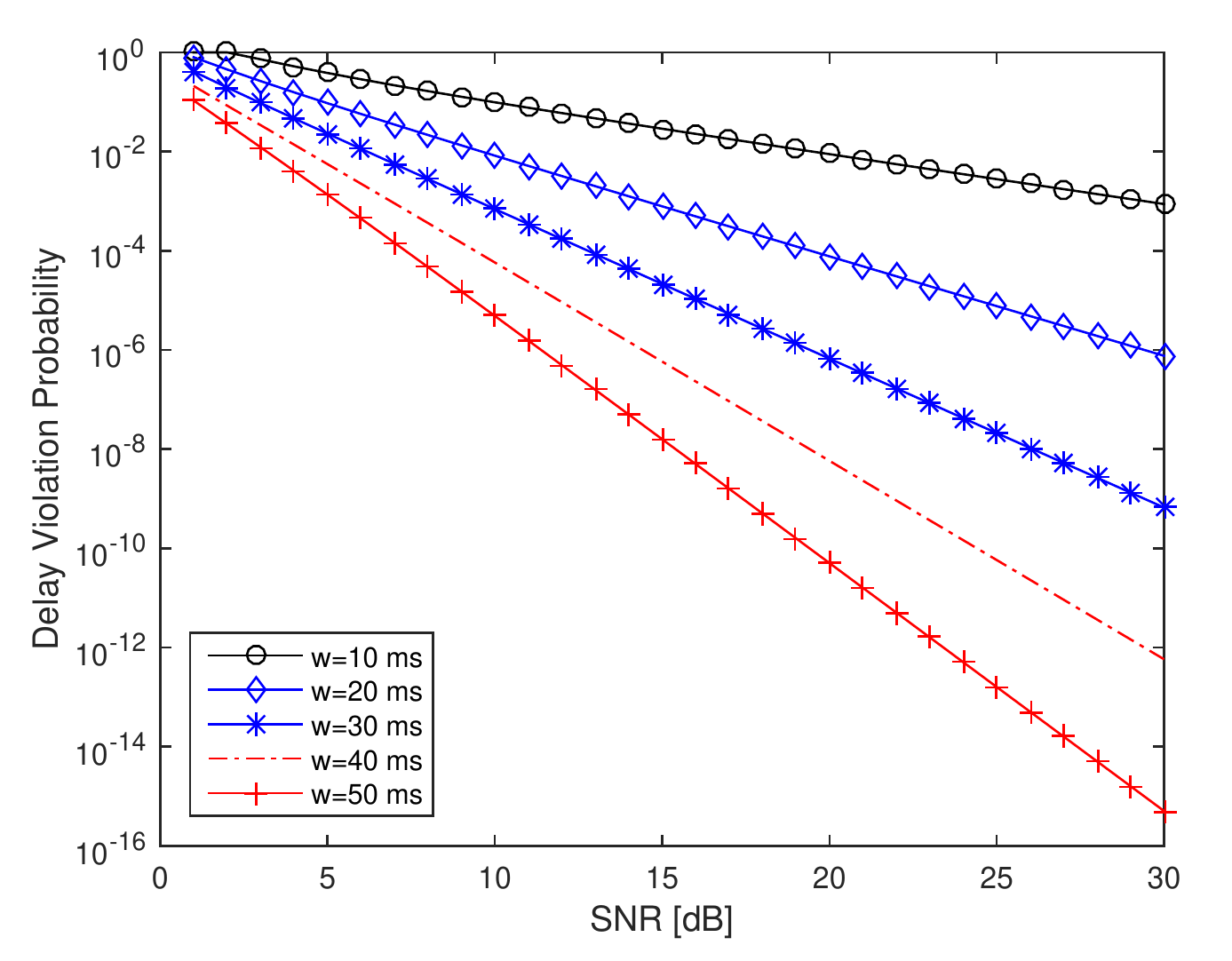} 
\caption{Delay violation probability depending on the average SNR computed for a single hop for various target delays and a packet size of 10 B.}
\label{fig:1hop}
\end{figure}
\begin{figure}
  \includegraphics[scale=0.6]{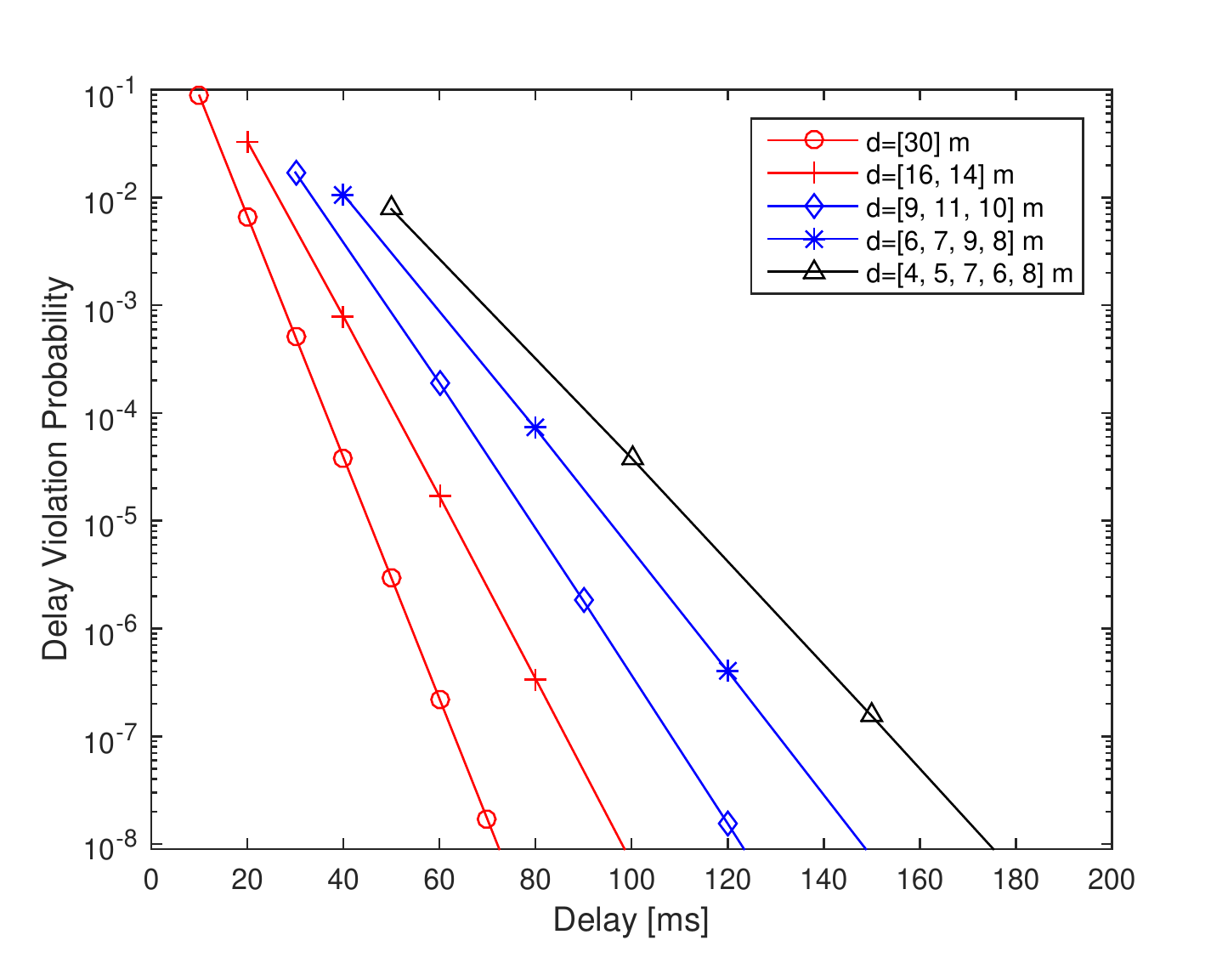} 
\caption{Analytical delay bound in a WirelessHART network vs. different target delays for several multi-hop paths. The distance between the source and the destination node in every scenario is 30 m and the total transmit power along the path is 4 dBm. The payload size is $r_a=10$ B.}
\label{fig:5hops}
\end{figure}
\begin{figure}[ht]
\centering
  \includegraphics[scale=0.6]{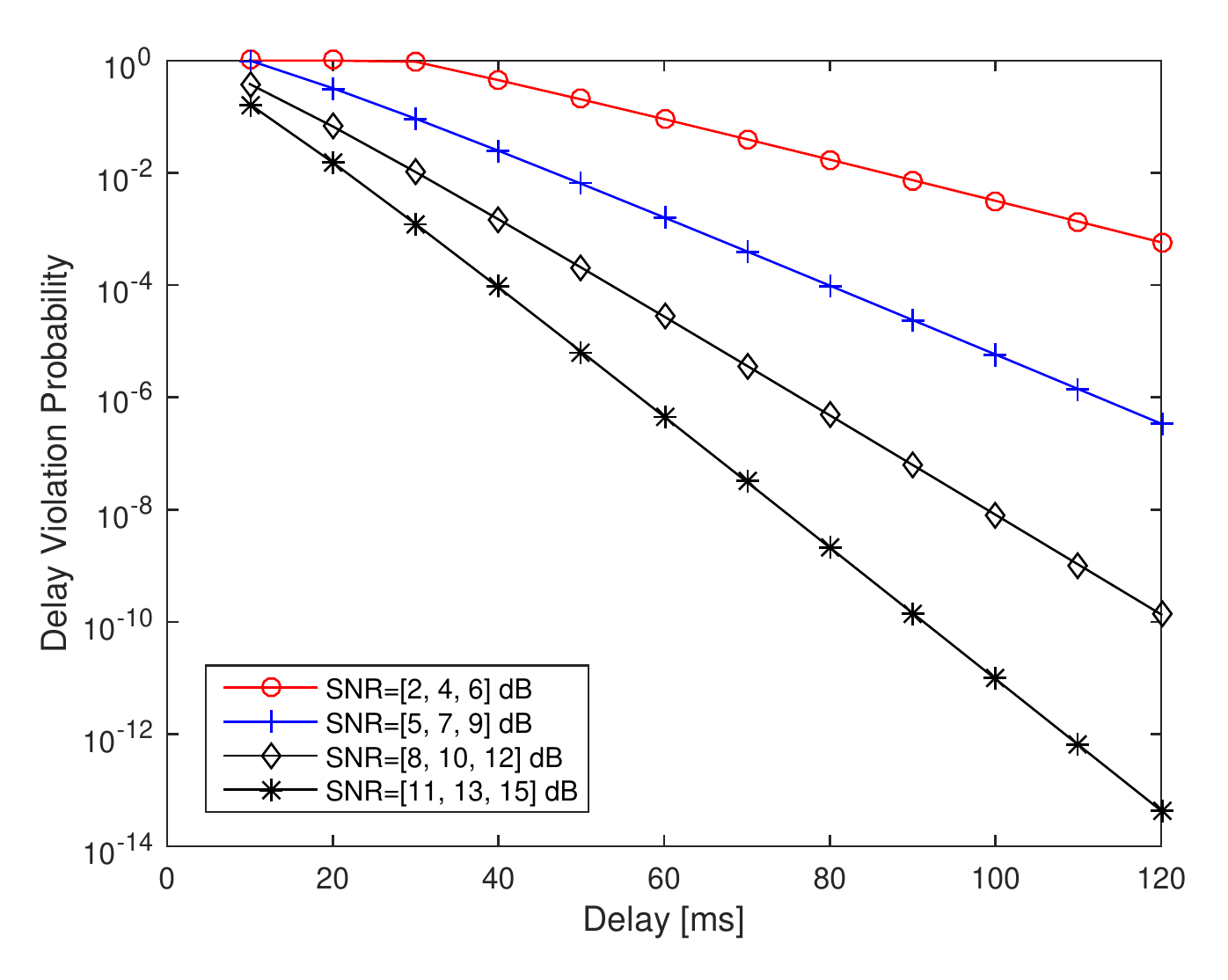} 
\caption{Doubling the SNR per link leads to two orders of magnitude lower delay bound.}
\label{fig:3dB}
\end{figure}

In Fig.~\ref{fig:diff_ra} - Fig.~\ref{fig:3dB} we focus solely on a numerical evaluation of the analytical results. Fig.~\ref{fig:diff_ra} shows how increasing the incoming payload size increases the delay violation probability for different target delays $\wepsilon=\{5,10,20\}$ superframes, which translates to $\{50,100,200\}$ ms in case of a single-hop communication. Obviously, using higher SNR on the link leads to lower violation probability for the same target delay and payload size. We notice that doubling the SNR on the link (which by constant noise level means roughly doubling the transmit power) results in a significant decrease of the violation probability for larger target delays ($w=200$ ms). This motivates us to look into ways of optimizing the SNR (usually by increasing the transmit power) on the link in order to improve the network performance for certain QoS requirements.
Similarly, Fig.~\ref{fig:1hop} shows how increasing the SNR in a single-hop communication can significantly decrease the delay violation probability, even for stricter QoS requirements, where the target delay is smaller than 50 ms. We notice that increasing the SNR can lead to violation probabilities in the order of $10^{-3}$ or $10^{-6}$ even for tighter delays ($w$=10 ms and $w$=20 ms, respectively). The payload size is $r_a=10$ B.

In Fig.~\ref{fig:5hops} we illustrate the end-to-end delay bound in case of 5 different path scenarios. In all cases the source and the destination node are 30 m apart. We create multi-hop paths by placing intermediate nodes and we equally distribute a total of 4 dBm transmit power among the nodes. Although the SNR of each link is increased while introducing additional links, adding more hops results with higher end-to-end delay bound. However, depending on the target delay and violation probability, multi-hop communication might not necessarily harm the QoS, but on the other hand, can reduce the total power consumption, which in turn will lead to a longer battery lifetime in case of battery-powered devices. This demonstrates again the strength of the framework, enabling a proper analysis prior network installation, leading to both performance guarantees and resource savings. 


Fig.~\ref{fig:3dB} illustrates the analytical delay bound for four different 3-hop path scenarios, obtained when doubling the SNR of the previous path. We notice that the gap between the corresponding delay violation probabilities increases as the target delay gets looser. Again, we focus on small payloads, typical for process automation and set their size at 10 B. 

\begin{figure}
\centering
  \includegraphics[scale=0.6]{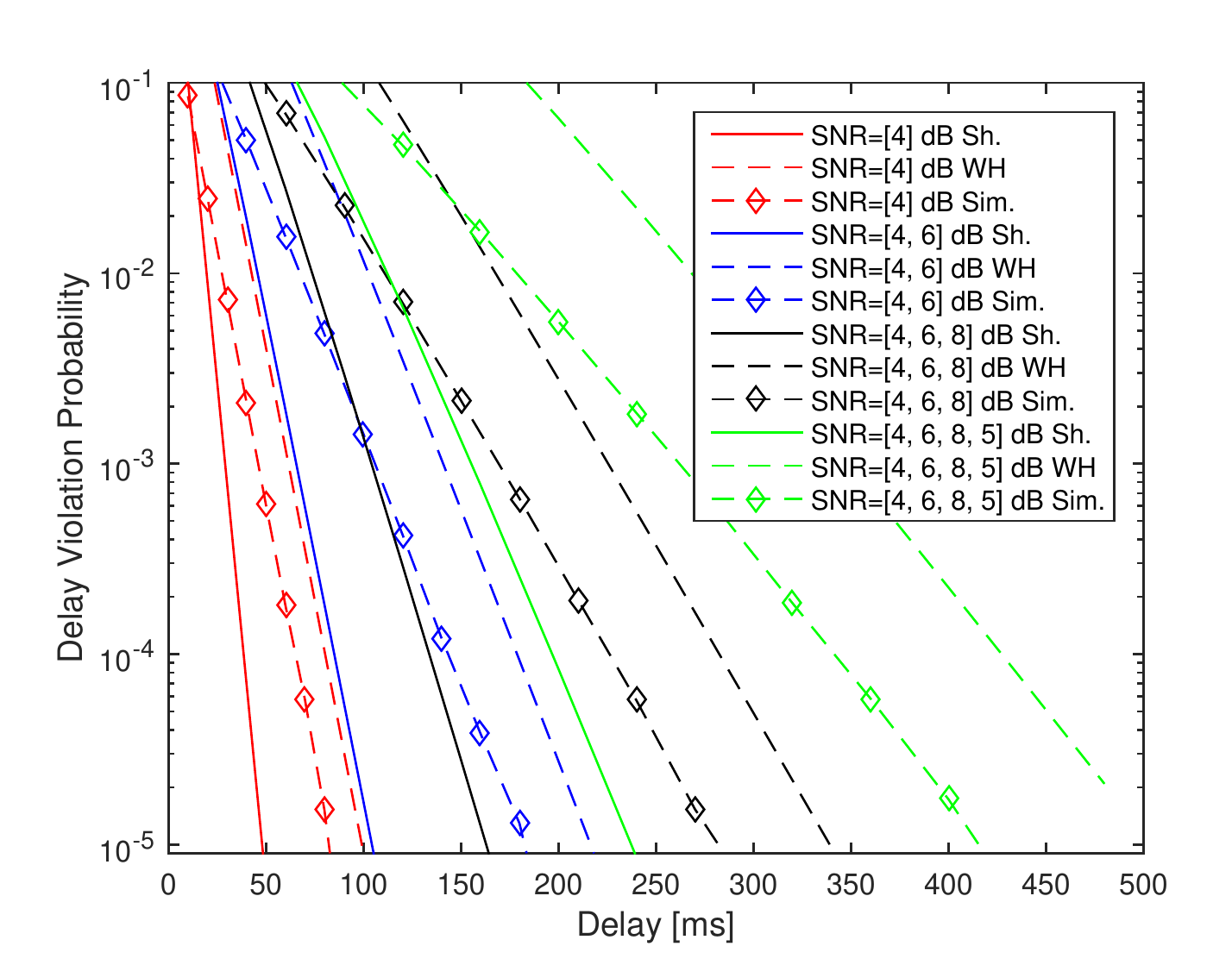} 
\caption{Delay bounds computed for cases of Shannon- and WirelessHART-based wireless channel capacity. The simulated delay violation probability is illustrated as well.}
\label{fig:two_bounds}
\end{figure}
As already mentioned in Sec.~\ref{sec:mot}, Shannon channel capacity is most frequently used in the research community for network analysis. Therefore, we are finally interested into the comparison of the delay bounds obtained when using the well known upper limit on the capacity versus the service description used in this paper. 
Since Shannon-based channel models represent a rather theoretical bound on the capacity and result in more optimistic delay guarantees, they are not the best choice for performance analysis of practical communication systems. We illustrate this in Fig.~\ref{fig:two_bounds}. The figure shows 
the analytical delay bounds obtained in case of service shaped according to the Shannon capacity (solid lines) on one hand and the IEEE 802.15.4 physical layer definition (dashed lines) on the other hand. We represent several multi-hop scenarios in a WirelessHART network. The instantaneous service of link $j$ in superframe $i$ according to Shannon is given by $s_{i,j}=C\log(1+\gamma_{i,j})$, where $C$ is the number of payload symbols that can be transmitted per time slot. Since IEEE 802.15.4 has a symbol rate of 62500 symbols/s, we take $C=625$ symbols during one time slot of 10 ms. The delay bound in such case was already defined in~\cite{alzubaidy} and also presented in Eq.(13) in~\cite{itc14}:
\begin{equation}
\label{eq:function_M}
  \K(s,-\wepsilon)=\frac{\left(e^{\nicefrac{1}{\gammabar_j}} \cdot {\gammabar_j}^{-s\mathcal{C}}\cdot \Gamma(1-s\mathcal{C},\frac{1}{\gammabar_j})\right)^{w}}{1-e^{r_a s}\cdot e^{\nicefrac{1}{\gammabar_j}} \cdot {\gammabar_j}^{-s\mathcal{C}}\cdot \Gamma(1-s\mathcal{C},\frac{1}{\gammabar_j})} \leq \varepsilon,
\end{equation}
\noindent where from Eq.~\eqref{eq:stability_cond} and for $\C=\nicefrac{C}{\log{2}}$ we obtain the stability condition $e^{r_a s}e^{\nicefrac{1}{\gammabar_j}}\gammabar_j^{-s\C}\Gamma(1-s\C,\frac{1}{\gammabar_j})<1$. We notice that the kernel computed according to Eq.~\eqref{eq:function_M} is not an upper bound on the simulated behaviour, since it overestimates the available channel capacity and therefore results with smaller delay violation probabilities. 
The provided service curve and delay bound given by Eq.~\eqref{eq:func_M} yield a more precise view of WirelessHART networks than the widely used Shannon capacity model, opening further paths for evaluation and design of industrial wireless networks.

\section{Conclusion}
\label{sec:conclusion}

In this work we present an analytical delay bound for wireless multi-hop networks, whose physical layer follows the definition of the IEEE 802.15.4 standard and the MAC layer incorporates time-slotted transmissions and frequency hopping. This makes our result applicable to different wireless industrial technologies, such as WirelessHART, ISA100.11a or TSCH-based networks. The proposed delay analysis is especially useful in the network design and flow admission process of wireless industrial networks, providing significant insights on their QoS-performance. We have validated the derived analytical results via simulations and have further discussed how modifying certain parameters, like e.g. the links' SNR, the hop number or the size of the incoming traffic can influence the delay violation probability. Moreover, the provided service curve can be used to determine other performance guarantees by means of stochastic network calculus, such as the backlog bound. The provided end-to-end delay bound characterized with its recursive nature enables extension of the presented results onto delay-aware routing algorithms for wireless industrial networks. Its dependance on the channel SNR opens further research potential in the area of QoS-aware power management, especially important for energy-limited sensor networks. We currently work towards combining these concepts. 

%
\bibliographystyle{abbrv}
\bibliography{bibl}  
%
%

\end{document}